\documentclass[twocolumn,prb,showpacs,preprintnumbers,amsmath,amssymb,floatfix]{revtex4}

\usepackage{graphicx}
\usepackage{dcolumn}
\usepackage{bm}
\usepackage{amssymb}
\usepackage{amsmath}

\begin{document}

\title{Time-Integrated Evidences for Superfluorescence from Magneto-plasmas in Semiconductor Quantum Wells}

\author{ Y. D. Jho$^{1}$}\email{jho@gist.ac.kr} \author{X. Wang$^{2}$, D.H. Reitze$^{2,3}$, J.
Kono$^{4}$, A.~A.~Belyanin$^{5}$, V.~V.~Kocharovsky$^{5,6}$,
Vl.~V.~Kocharovsky$^{6}$, and G.~S.~Solomon$^{7}$}

\affiliation{$^1$Department of Information and Communications,
Gwangju Institute of Science and Technology, Gwangju 500-712,
Republic of Korea\\ $^2$Department of Physics, University of
Florida, Gainesville, Florida 32611, USA\\ $^3$National High
Magnetic Field Laboratory, Florida State
University, Tallahassee, FL 32310, USA\\
$^4$Department of Electrical and Computer Engineering, Rice
University, Houston, Texas 77005, USA\\
$^5$Department of Physics,
Texas A\&M University, College Station, Texas 77843, USA\\
$^6$Institute of Applied Physics, Russian Academy of Sciences,
603950 Nizhny Novgorod, Russia\\ $^7$Solid-State Laboratories,
Stanford University, Stanford, California 94305, USA}

\begin{abstract}
We present a comprehensive series of investigations of light
emission from semiconductor multiple quantum wells in strong
magnetic fields excited by intense femtosecond laser pulses. An
analysis of the dependence of the inter-Landau level (LL) emission
on magnetic field strength and laser fluence as well as the
spatial and statistical characteristics of the emitted light
indicate that the initial photo-excited electron-hole pairs
spontaneously form a macroscopic coherent state upon relaxation
into the LLs, followed by the emission of a a superfluorescent
burst of radiation. Our results are in good agreement with the
predicted characteristics for superfluorescent emission from
semiconductor quantum wells. We also investigate the effects of
spot size, temperature, excitation geometry, and excitation
pulse-width on the emission properties.
\end{abstract}

\pacs{78.20.Ls, 78.55.-m, 78.67.-n} \maketitle

\section{Introduction}

Investigations of coherent phenomena in bulk and quantum-confined
semiconductor systems have been made possible in the last fifteen
years through the use of ultrafast lasers, which create and probe
quantum coherence between electrons and holes on time scales
faster than inherent phase-breaking (decoherence) times in these
systems, typically 10-100 fs for free carriers to a few
picoseconds for excitons. During the coherent temporal regime, the
photo-excited carriers retain a well-defined phase relationship
with the excitation field, and the induced carrier polarization
can be followed as the polarization decays, thus probing
fundamental quantum decoherence and scattering. A vast array of
ultrafast coherent dynamics in semiconductors has been explored.
\cite{Shah} More specifically, investigations on coherent quantum
optics in condensed matter and mesoscopic systems are actively
being pursued~\cite{Yamamoto}. In particular, ultrafast
semiconductor analogs of quantum optical phenomena such as Rabi
flopping~\cite{Cundiff} and superradiance~\cite{XSR} have been
investigated.

An equally important but much less studied phenomenon in
semiconductors and nanostructures is the formation of spontaneous
macroscopic quantum coherence. Under specific physical conditions,
an ensemble of interacting particles can \emph{spontaneously}
establish quantum coherence. The spontaneous formation of a
coherent quantum ensemble is fundamentally different from
laser-driven coherence. Typically, such transitions occur at a
critical temperature and/or density at which the particle
interactions energetically favor a single N-body wave function.
Bose-Einstein condensation (BEC) ~\cite{Weiman}, superconductivity
~\cite{DeGennes66}, and superfluidity ~\cite{superflu} are typical
examples.  While the observation of spontaneous quantum coherence
in semiconductors has long been sought after, investigations have
been heretofore hampered by the strong electronic interactions and
concomitant rapid dephasing times in these systems.  Nevertheless,
investigations pursuing spontaneous coherence are important
because they can reveal much about the underlying fundamental
quantum mechanical interactions of N-particle systems. More
importantly, they can lead to routes for achieving such sought
after goals as BEC in solids. As an example, evidence for a new
kind of macroscopic ordered exciton state has been seen in
indirect (Type II) quantum wells at sub-Kelvin temperatures
excited by ultrafast laser pulses.\cite{butov-nature}

Superfluorescence (SF) is a well known example of spontaneously
created macroscopic quantum coherence in quantum optics and is a
particularly good candidate for these types of investigations. In
SF, an incoherently prepared system of inverted two-level atoms
develops coherent large amplitude oscillations of the optical
polarization starting from the level of quantum fluctuations
through the self-phasing process mediated by exchange of photons
~\cite{Siegman, Zheleznyakov89, Bonifacio75}. The resultant
macroscopic optical polarization decays superradiantly
~\cite{Dicke54,Rehler71} producing a highly directional burst of
coherent radiation.  Until now, SF had been observed only in
atomic gases ~\cite{Skribanowitz73,Gibbs77} and rarefied
impurities in crystals ~\cite{Florian82,Zinoviev83,Malcuit87}.
Observations of SF emissions from high density systems such as
semiconductor electron-hole plasmas have heretofore proven
difficult.

We recently reported on the first observation of SF from a dense
photo-excited electron-hole plasma in InGaAs/GaAs multiple quantum
wells in strong perpendicular magnetic fields ~\cite{JhoSF}. Using
intense femtosecond laser pulses to create a dense electron-hole
plasma in InGaAs/GaAs multiple quantum wells, combined with a
strong perpendicular magnetic field to laterally confine electrons
and holes and increase the density of states through Landau
quantization, we observed the emission of SF bursts from
inter-Landau level (LL) recombination as evidenced from the
resulting spectral, spatial, and statistical characteristics of
the luminescence. Our investigations revealed a transition from a
spontaneous emission (single particle recombination) regime at low
carrier densities and magnetic fields through an intermediate (and
deterministic) amplified spontaneous emission (ASE) regime to a
stochastically emitting SF regime at the highest densities and
magnetic fields.  In this paper, we present an expanded series of
investigations into the time-integrated characteristics of SF
emission. We examine how SF is influenced by magnetic field, laser
pump pulse fluence, spot size, temperature, and the excitation
focal geometry. We also investigate how sublevel mixing influences
the emission strength. Finally, we demonstrate that the direction
of emission can be efficiently controlled by tailoring the shape
of the excitation geometry.

The paper is organized as follows.  In Section ~\ref{background},
we present the theoretical basis for SF from quantized
electron-hole plasmas, present key physical parameters, and
discuss the different density- and field-dependent emission
regimes. We briefly describe the experimental procedures and
details in Section ~\ref{exp}. Our experimental results are given
in Section ~\ref{results}, where we present a comprehensive set of
magneto-optical photoluminescence spectra under a variety of
relevant conditions. Specifically, we examine the spectral
dependence on magnetic field, excitation fluence and duration,
excitation geometry, and temperature as well as the spatial
characteristics of the emission in different regimes.  In
addition, analyses of the field scaling and temperature dependence
are given. We conclude in Section ~\ref{conc}.

\section{Theoretical Background}\label{background}

The physics of SF in atomic systems has been studied for more than
thirty years; see. e.g., the reviews
~\cite{Bonifacio75,Gross82,Zheleznyakov89,Andreev93,Benedict96}
and references therein.  An incoherently prepared ensemble of
excited two level dipoles interacts via the exchange of
spontaneously emitted photons. At sufficiently high densities $N$
of inverted atoms, this exchange leads to an effective mutual
phasing of the atomic dipole oscillators. As a result, a
macroscopic polarization with an amplitude $\propto N$
spontaneously forms arising from quantum fluctuations on a time
scale $t_d$. The resulting polarization decays by emitting a
coherent pulse of collective (but spontaneous) radiation with
duration $t_p$ scaling as $1/N$ in the simplest case, so that the
emitted intensity $I_{SF} \sim N/t_p \sim N^2$ exceeds by many
orders of magnitude the intensity $I_{SE}$ of spontaneous emission
from the same number of incoherent atoms ($I_{SE} \sim N/T_1 \sim
N$). The  SF pulse duration $t_p$ and the delay time $t_d$ are
\emph{less} than the times of incoherent spontaneous emission and
phase relaxation in the medium:  $t_p, t_d < T_1, T_2$. This is a
distinctive feature of the phenomenon. In addition, SF is
fundamentally a stochastic process: the optical polarization and
the electromagnetic field grow from initially incoherent quantum
noise to a macroscopic level. Thus, SF is \emph{intrinsically
random}: for identical preparation conditions, initial microscopic
fluctuations get exponentially amplified and may result in the
macroscopic pulse-to-pulse fluctuations in the delay time $t_d$,
electric field polarization, and the direction of the emitted
pulse.

We note that there is some terminological confusion in the
literature between the two types of cooperative emission
processes, superradiance and superfluorescence. Superradiance
results when the formation of coherent polarization is driven by
an external pump laser field, whereas a superfluorescent
macroscopic polarization develops spontaneously due to an
intrinsic instability in a system of initially incoherent quantum
dipole oscillators. Superradiance is generally easier to observe,
and has been previously observed in semiconductors \cite{SR}.

The key to achieving SF is to provide high enough spatial and
spectral density of inverted dipole oscillators (atoms, excitons,
electron-hole pairs etc.), such that the growth rate of the
polarization exceeds the dephasing rate ($1/T_2$). In
semiconductors, the rate of the optical polarization decay is
typically 1-10 ps$^{-1}$ or greater, whereas the maximum density
of electron-hole pairs within the bandwidth $\Delta \omega$ of the
SF pulse is limited by the density of states $\rho(E)$: $N_{max}
\sim \rho(E) \hbar \Delta \omega$. It is this combination of a
limited gain and fast decoherence that makes it difficult to
achieve the threshold for SF ~\cite{BKK92,Belyanin97}. One
possible way to overcome these limitations is to use quantum wells
placed in strong magnetic fields. The combination of reduced
dimensionality and the magnetic field fully quantizes the
semiconductor into an atomic-like system with a series of Landau
levels (LLs), thus strongly increasing the density of states (DOS)
~\cite{JhoSF}. In addition, a complete quantum confinement and
filling of all available electron states at the lowest LLs are
expected to suppress the dephasing rate of the optical
polarization. In our experiments, the initial pumping conditions
are chosen so as to ensure the growth of coherence from an
incoherent state by exciting the initial e-h plasma high in the
bands with an excess energy of 150 meV above the GaAs barrier band
gap. The energy difference between the excitation energy and the
0-0 LL in our QWs ~\cite{MacDonald} is greater than 270 meV,
requiring many scattering events to relax into the QW LLs. The
resulting e-h plasma in the individual LLs is thus initially
\emph{completely incoherent}.

\subsection{Semiclassical Model of Superfluorescence}

To understand the threshold behavior and main features of SF, we
will use a coupled density matrix and Maxwell equations in the
semiclassical approximation, i.e. assuming a classical
electromagnetic field ~\cite{qed}. Note that although the optical
polarization of an ensemble of quantum oscillators as a whole
emerges spontaneously, individual atoms interact with common
electromagnetic mode through \emph{ stimulated} emission and
absorption processes. As the field increases exponentially,
semiclassical treatment becomes applicable early in the process
and determines its main features. This approach captures the main
features of both exponential amplification and subsequent SF pulse
formation in various systems ~\cite{Zheleznyakov89}.

In our experiment, electrons and holes are created by a pumping
pulse high in the bands and then quickly lose energy, ending up in
the ground subband of QWs before SF starts to develop.  We do not
consider the initial kinetics of rapid energy and momentum
relaxation of carriers with a time scale of 10-500 fs, since these
hot carriers do not contribute to SF. We concentrate on carriers
that have already reached a quasi-equilibrium degenerate
distribution in the lowest LL. We furthermore consider radiation
and optical polarization resulting from stimulated optical
transitions between electrons and holes on a given LL with the
same quantum number. Choosing $z$ as the growth direction, we can
write the electric field for $y$-polarized electromagnetic waves
propagating in $x$-direction as $E_y = (1/2) E_t(y,z) A(x,t)
\exp(ikx-i\omega t)$, where $E_t$ is a transverse distribution of
the field. It satisfies a 2D Helmholtz equation that determines
transverse modes of the structure. For the slowly varying
amplitude field $A$ of a given mode and the amplitude polarization
$P$ one can obtain
\begin{equation} \label{rate1}
\frac{\partial A}{\partial x} + \frac{\mu}{c} \frac{\partial
A}{\partial t} + \alpha A = \frac{2 \pi i \omega}{\mu c} \Gamma P,
\end{equation}
\begin{equation} \label{rate2}
\frac{\partial \rho_0}{\partial t} = - (\gamma + i \Delta) \rho_0
+ i \Omega_R (n_e + n_h -1),
\end{equation}
\begin{equation} \label{rate3}
\frac{\partial n_e}{\partial t} = {\rm Im}[\Omega_R^* \rho_0] +
R_e - n_e/T_e(n_e-f_e)/\tau_e,
\end{equation}
\begin{equation} \label{rate4}
\frac{\partial n_h}{\partial t} = {\rm Im}[\Omega_R^* \rho_0] +
R_h - n_h/T_h(n_h-f_h)/\tau_h.
\end{equation}
In Eq. (1), $\alpha = 2 \pi \sigma/(\mu c)$ is the losses of a
given electromagnetic mode with the refractive index $\mu$, the
non-resonant Ohmic conductivity of a sample at the optical
frequency $\sigma(\omega)$, and  the velocity of light in vacuum
$c$. $\Gamma$ is an overlap of the transverse distribution of the
electromagnetic field with the MQW region. Here the amplitude of
the polarization $P$ is related to the amplitude of the
off-diagonal element of the density matrix $\rho = \rho_0
e^{-i\omega t}$ for an electron and hole state with occupation
numbers $n_e$ and $n_h$ and the dipole moment $d$ of the optical
transition as $P = (1/V) \sum_i d_i \rho_{0i}$, where the
summation is over all electron-hole pairs on a given LL. Total
volume densities are determined by a similar summation of $n_e$
and $n_h$ over the volume $V$ of a structure. In Eq. (2), $\Delta
= \omega_0 - \omega$ is the detuning of the frequency of the
electromagnetic mode from the optical transition frequency
$\omega_0$ and $\gamma$ is the transverse decay rate. $R_{e,h}$ is
the rate of influx of carriers from upper LLs and barriers;
$T_{e,h}$ are total lifetimes of a given state due to radiative
and non-radiative processes. The last terms in Eqs. (3) and (4)
describe fast relaxation to quasi-equilibrium distributions
$f_{e,h}$ with relevant time constant $\tau_{e,h}$ and $\Omega_R$
is the Rabi frequency of the electromagnetic mode ~\cite{rabi}.

The optical transition frequency $\omega_0$, which determines
$\Delta$, is renormalized by screened Coulomb interaction with
other carriers ~\cite{HK94,WH98}. This effect leads to a redshift
of the transition energy from its single-particle value $E_s = E_g
+ E_{e1} + E_{hh1} + (n+ 1/2)\hbar eB/(m_r c)$, where $n$ is the
number of a given LL, $E_{e1}$ and $E_{hh1}$ are the edges of the
ground subbands of electrons and heavy holes, and $1/m_r = 1/m_e +
1/m_{hh}$. For our structure $E_s \simeq 1.345 + 0.0023 B(T)$ eV,
where the magnetic field $B$ is expressed in Tesla. By far the
most important effect is the resonant interaction with
electron-hole pairs on a given LL which defines both the growth
rate of the instability and the polariton shift of the refractive
index $\mu$.  In this regard, $\Gamma$ is a crucial parameter in
our gain-guided sample which strongly depends on the polariton
contribution to the refractive index coming from excited carriers;
see below.

The discrimination between SF and incoherent recombination regimes
 is convenient to formulate in terms of a linear initial-value
problem, assuming initial quasi-equilibrium degenerate populations
of carriers created by pumping and subsequent relaxation. After
finding the linear susceptibility $\chi(\omega)$ from Eq. (2) one
can solve the dispersion relation following from Eq. (1) and find
the complex frequency $\omega(k)$ as a function of a real wave
number $k$. Note that slow-varying amplitude approximation implies
that $|\omega|$ and $k$ are close to resonance frequency
$\omega_0$ and wave number $\mu \omega_0/c$. The instability, or
exponential growth of the field and polarization with time,
develops when Im$[\omega] > 0$. Obviously, for instability of any
kind one needs positive population inversion $\Delta N \equiv
n_e+n_h-1 >0 $ with respect to stimulated recombination vs.
absorption processes at the photon frequency.

Consider first an idealized situation when all scattering rates
and transition frequencies are the same for all electron-hole
pairs on a given LL, similarly to a homogeneously broadened active
medium. Note that each spin-split LL has a degeneracy of $N_{2D} =
eB/(2\pi\hbar c) \simeq 2.4\times 10^{10} B(T)$ cm$^{-2}$, and
total population inversion $\Delta N > 0$ can be defined as the
difference between occupied and unoccupied states per unit area.
In this case the dispersion relation $k^2 = \omega^2/c^2[\mu^2+ 4
\pi i \sigma/\omega + 4\pi \chi(\omega)]$ has two solutions
$\omega(k)$ -- eigen modes that are sometimes called
electromagnetic and polariton branches.

\subsection{Emission Regimes and Growth Rates}

The key parameter governing the instability of these modes and
different regimes of recombination is the cooperative frequency
$\omega_c$ that determines the coupling strength between the field
and the optical polarization as introduced in previous
studies\cite{Belyanin97,Belyanin98,Belyanin03}
\begin{equation} \label{coop}
\omega_c = \sqrt{\frac{16 \pi^2 d^2 \Delta N \Gamma c}{\hbar \mu^2
\lambda L_{QW}}}.
\end{equation}

Here $L_{QW}$ is the total width of the QWs. When the inversion
$\Delta N$ is small or negative, field oscillations decay with
time and one can have only spontaneous recombination emission with
power $\propto N_e/t_{sp}$ and characteristic timescale $t_{sp}
\sim 1$ ns, where $N_e$ can be presumed to be the same with the
degeneracy in a LL in our experimental condition. The inversion
can be increased by increasing the fluence of a pumping pulse and
the magnetic field strength. With increasing inversion and optical
confinement the modal gain may become higher than losses and
oscillations of the field and polarization will grow with time.
There are two basic regimes of the instability, depending on the
ratio between the values of $\omega_c$ and incoherent relaxation
times. They correspond to the instability of one of the two eigen
modes existing in the medium. For low inversion density and fast
decoherence, when $\omega_c \ll \gamma = 1/T_2$, one can have an
amplified spontaneous emission (ASE) provided photon losses
$\alpha $ are low enough. This is a regime of a one-pass
amplifier, corresponding to the instability of the electromagnetic
branch. Its growth rate
\begin{equation} \label{lase}
Im[\omega] = g_{ASE} \approx \frac{\omega_c^2 \gamma}{4(\gamma^2 +
(k-k_0)^2)} -1/T_E \ll \gamma
\end{equation}
 is much slower than the dephasing rate $\gamma$, where $T_E$ is the photon transit time through the active region.

When the gain increases to the value such that $g_{ASE}L \mu/c \gg
1$, amplification proceeds in the saturated regime which is
sometimes called superluminescence (SL). Here $L$ is the length of
an active medium in the propagation direction. The duration of the
SL pulse in a saturated amplifier is $t_{SL} \sim L \mu/c$.
~\cite{Benedict96}

If there is feedback for any kind of modes existing in a sample,
e.g., Fabry-Perot cavity modes formed by reflections from the
edges, whispering gallery modes etc., then a system can lase on
these modes. The growth rate of lasing modes will be defined by
the same expression (~\ref{lase}) with the optical confinement
$\Gamma$ and photon lifetime $T_E$ defined for a given laser mode.
One-pass amplification and lasing compete with each other and the
process with a higher growth rate dominates. We argue below that
there is no feedback and therefore no lasing for our samples, but
it is important to stress here that \emph{all} timescales for
\emph{all} of the above processes of ASE, SL, or lasing are longer
than the dephasing time $T_2$. Under such conditions the polariton
mode is strongly damped with the decay rate $\gamma$ and one can
adiabatically eliminate optical polarization from
Eqs.~(~\ref{rate1})-(~\ref{rate4}) assuming $\partial
\rho_0/\partial t = 0$ and $|\Delta| \ll \gamma$. The growth rate
(~\ref{lase}) has nothing to do with coherent dynamics of
individual dipoles and originates from amplification of
electromagnetic waves adiabatically followed by polarization.

Cooperative recombination regime (SF) results from the spontaneous
formation of a large-amplitude coherent macroscopic polarization
from initially incoherent oscillations of individual electron-hole
dipole moments. It requires that $\omega_c
> 2\gamma$, or $\omega_c > 2(\gamma \gamma^*)^{1/2}$ for a system with a large inhomogeneous
broadening $\gamma^* \gg \gamma$ of frequencies $\omega_0$ of
electron-hole dipole oscillators
~\cite{Zheleznyakov89,Belyanin97,Belyanin98}.

When the field dissipation rate is very high, $\gamma < \omega_c
/2 < 1/ T_E$, cooperative emission develops with a growth rate
$g_{SF} \approx \omega_c^2 T_E/4 \propto \Delta N$. This
instability of the polariton branch occurs \emph{because of} Ohmic
dissipation (the dissipative instability). However, it is not
relevant for our samples that are characterized by a low field
dissipation rate: $\omega_c /2
> \gamma > 1/ T_E$.  In this case the growth rate saturates at its maximum value
$g_{SF} \approx \omega_c/2 \propto (\Delta N)^{1/2} $. Note that
in all cases the SF growth rate is faster than the phase
relaxation rate $\gamma$. This ensures maximally coherent nature
of the process. In a sense, SF establishes an absolute upper limit
on the rate with which an ensemble of initially incoherent
inverted oscillators can radiate their stored energy.

The dynamics of SF is also noteworthy and very different from ASE
or lasing. First, SF develops with a broad spectral bandwidth $~
g_{SF}> \gamma$ and thus cannot be described by usual rate
equations based on adiabatic elimination of the polarization.
Second, after the degenerate (inverted) population of carriers is
established on a LL, the SF pulse is emitted with the delay time
$t_d \sim (1/g_{SF}) {\rm ln}[I_{SF}/I_{0}]$ which is
logarithmically larger than the inverse growth rate, where the
logarithm factor of order 10-20 is due to the exponential growth
of the field from the quantum noise level $I_{0}$ to the peak
intensity $I_{SF}$.  Third, the pulse duration $\tau_{SF} \sim
2/g_{SF}$ decreases with $\Delta N$ and therefore the pulse
intensity $I_{SF} \sim  N_e/\tau_{SF}$ scales superlinearly
$\propto N_e^{3/2} \propto B^{3/2}$ with electron density or the
magnetic field, assuming that $\Delta N \sim N_e$. The coherence
length over which an individual pulse is formed is given by $L_c
\sim ct_d/\mu$, which is a logarithmic factor of order 10 longer
than the exponential amplification length $\sim c/(\mu g_{SF})$.
Thus the volume of the active medium in which all electron-hole
dipoles become phased and contribute coherently into an
exponential amplification of a given electromagnetic mode is given
by $\sim \Gamma \lambda^2 c/(\mu g_{SF})$. If we multiply it by
the volume density of inverted oscillators $\Delta N/L_{QW}$ and
by the spontaneous recombination rate of an individual e-h pair
given by $1/t_{sp} = 32 \pi^3 d^2 \mu/(\hbar \lambda^3)$, we
obtain the maximum SF growth rate $\omega_c/2$ within a factor of
2.

The numerical value of the growth rate depends crucially on the
optical confinement factor $\Gamma$. The contrast in the
background refractive index $\Delta \mu \simeq 0.02$ between the
MQW layer and GaAs is too low to provide wave guiding in a highly
asymmetric waveguide formed by air on one side and GaAs on the
other side of the MQWs. In the absence of a gain-induced change in
the refractive index the modes are spread over the whole thickness
of the sample and the overlap factor $\Gamma$ is only $\sim
10^{-4}$ which rules out \emph{all} stimulated amplification
regimes, i.e., both SF and ASE. Numerical simulations show that
the minimum contrast needed for wave guiding is $\Delta \mu =
0.045$. The confinement factor $\Gamma$ is ~ 0.1 when $\Delta \mu
= 0.05$ and reaches 0.4 with subsequent increase of the index
contrast to 0.09 when $B = 25$ T. This affects the dependence of
the ASE or SF growth rate and of the peak pulse power on the
electron density or the magnetic field. Obviously the dependence
becomes stronger, which can be directly observed in time-resolved
measurements.

\subsection{Gain Guiding}

In an inverted medium the polariton contribution to the background
refractive index is positive on the blue side of the transition
frequency $\omega_0$, with the peak index change $\delta \mu
\simeq \pi d^2 \Delta N /(\hbar \mu L_{QW} \gamma)$ reached at a
detuning $\omega ?\omega_0 \simeq \gamma$. For a high pumping
fluence we can assume that all states are occupied and $\Delta N =
2N_{2D}$; then for $T_2 = 1/\gamma = 250$ fs $\delta \mu$ reaches
the wave guiding threshold of 0.03 when $ B \simeq 12$ T. The
latter value corresponds to the experimentally observed appearance
of the narrow stimulated emission peak on the blue side of the
broad spontaneous emission spectrum; see Fig. 1.

Thus the polariton-supported wave guiding provides natural
explanation of the observed blueshifted peak. If we adopt this
explanation, then the observed stimulated emission threshold can
be used to determine $\gamma$. The number we obtained supports the
suggestion that the dephasing time for quantum-confined degenerate
carriers on the LLs before the formation of the emission pulse can
be as long as several hundreds fs. Unfortunately, this value of
$\gamma$ is difficult to verify independently, e.g. from the width
of the spontaneous emission spectrum because $\gamma$ may change
significantly during one emission cycle due to varying density of
carriers, and we collected only time-integrated data. Also, there
could be some inhomogeneous broadening of the spontaneous emission
due to sample inhomogeneity within the pumped spot. However, the
width of the narrow stimulated emission peak does provide an
additional information on the values of $\gamma$ and the gain,
which is consistent with the above estimate, as discussed below.

At the onset of wave guiding the stimulated recombination is still
in the ASE regime, as discussed in more detail in Sec. 3A. The
transition to SF occurs at $B \geq 20$ T, when the net growth rate
$\omega_c/2 - \gamma/2 \simeq 8\times 10^{12}$ s$^{-1}$ and
$2N_{2D} \simeq 10^{12}$ cm$^{-2}$. The coherence length over
which a SF pulse is formed is $L_c = c t_d/\mu \sim c/(\mu g_{SF})
{\rm ln}[I_{SF}/I_{0}]$. The ratio $I_{SF}/I_{0}$ is of the order
of the total number $N_t$ of e-h pairs contributing to a given
pulse in the volume occupied by an electromagnetic mode, which we
estimate as $N_t \sim 10^7$. Assuming $g_{SF} \sim \gamma \sim 4
\times 10^{12}$ s$^{-1}$, we obtain $L_c \sim 0.4$ mm. For
observations of SF it is optimal to create an inverted (pumped)
spot of this size. At much shorter lengths a SF pulse does not
have time/distance to develop, while at longer distances $L \gg
L_c$ independent SF pulses will be formed at segments of length
$L_c$, creating a noisy output over the time interval $\sim
L\mu/c$. Note that in a sample of the optimal pumped length $L_c$
the shortest ASE pulse in a high gain regime has duration $\sim
L_c \mu/c \sim 5$ ps, whereas a SF pulse is $\leq 1$ ps.

After we defined all relevant spatial and temporal scales, let us
turn to the experiments and their interpretation.


\section{Experimental Methods}\label{exp}

\begin{figure}\begin{center}
\includegraphics[scale=0.4,trim=50 100 100
0,angle=0]{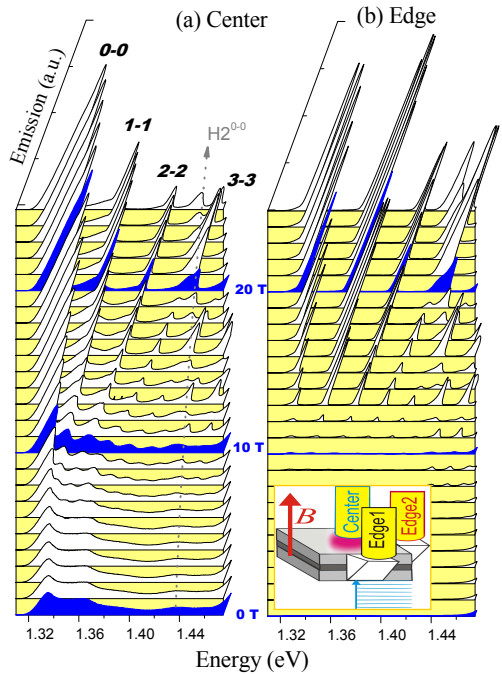}\end{center} \caption{(Color Online) Emission
spectra as a function of magnetic field (a) for emission through
sample above the excitation spot along the laser propagation
direction and (b) for in-plane emission in an edge collection
geometry. The inset of (b) shows the experimental schematic
showing excitation pulses from the bottom and the collection
fibers from the top.  The pump fluence was 0.62 mJ/cm$^2$.  The
gray dotted line denotes the position of the 0-0 LL of the e2-HH1
state.} \label{PL}
\end{figure}

Our experiments were performed in the new ``Fast Optics" Facility
at the National High Magnetic Experiments in Tallahassee, FL
~\cite{fastoptics} using a 31 T DC Bitter magnet.
In$_{0.2}$Ga$_{0.8}$As/GaAs quantum well samples, consisting of 15
layers of 8 nm thick QWs separated by 15 nm GaAs barriers grown by
molecular beam epitaxy, were used as samples. The magnetic field
dependent low-power CW PL and absorption spectra of these samples
are thoroughly described in Ref. ~\onlinecite{Jho}. We used a 1
kHz repetition rate 775 nm or 800 nm Ti:Sapphire chirped pulse
regenerative amplifier system to excite the sample. Both
pulsewidth and fluence were varied to investigate scaling laws.
For reference, a pulse duration of 150 fs and pump fluence of
$\sim 0.01$ mJ/cm$^2$ generates a carrier density of $\sim$
10$^{12}$ cm$^{-2}$ in our samples.

The femtosecond excitation beam was delivered through free space
into the magnet, and the plane of the QW was perpendicular to $B$
with field up to 30 T. The emitted photoluminescence (PL) was
collected using multimode optical fibers from the opposite face
(center fiber) and cleaved edges of the sample (in the plane of
the QW) using right-angle micro-prisms coupled to multi-mode
optical fibers and analyzed with a high resolution MacPherson
grating spectrometer equipped with a charge-coupled device (CCD)
detector. The collection area of the prisms was 1x1 mm$^2$, and
the computed acceptance angle based in geometric considerations
was $\sim 40^{\circ}$.  Most of the spectra were collected by
averaging the emission from approximately 1000 pulses; single
pulse excitation using an external Pockels cell was used for
excitation in experiments to probe the statistics of the
directionality of the emission (Section \textbf{IV C}).
Temperature-dependent spectra were collected from 10 K to 130 K to
determine how dephasing effects influence the emission process.
The pump spot size was varied in diameter as well as aspect ratio
to examine the characteristics of the emission relative to the
predicted coherence length of the electron-hole plasma.

\section{Experimental Results and Further Discussions}\label{results}

\subsection{Dependence of Photoluminescence on Magnetic Field and Laser Fluence}

Figure \ref{PL} displays emission spectra as a function of
magnetic field $B$ for a pump fluence of $\sim 0.62$ mJ/cm$^2$ (a)
from the opposite side of the sample above the pump spot (denoted
by `center'; see inset) and (b) at the sample edge perpendicular
to the excitation direction (`edge'). The spectra in Fig.
\ref{PL}(a,b) both show well defined and broad ($\Delta E\sim$ 9
meV) LL states as reported in previous studies
\cite{Butov,Potemski}.  However, at a threshold field $B$ = 13 T,
narrow spectral features ($\Delta E \sim$ 2 meV) emerge from
high-energy side of LL peaks. The narrow features grow rapidly and
become dominant at higher $B$. Although clearly visible in both
the center and edge geometries, the narrow emission features are
generated with higher efficiency in the edge geometry; the
integrated emission strength of the sharp feature in the 0-0 LL at
25 T is approximately 70 times stronger than the broad spontaneous
emission.

\begin{figure}
\includegraphics[scale=0.35,trim=20 20 50
20,angle=-90] {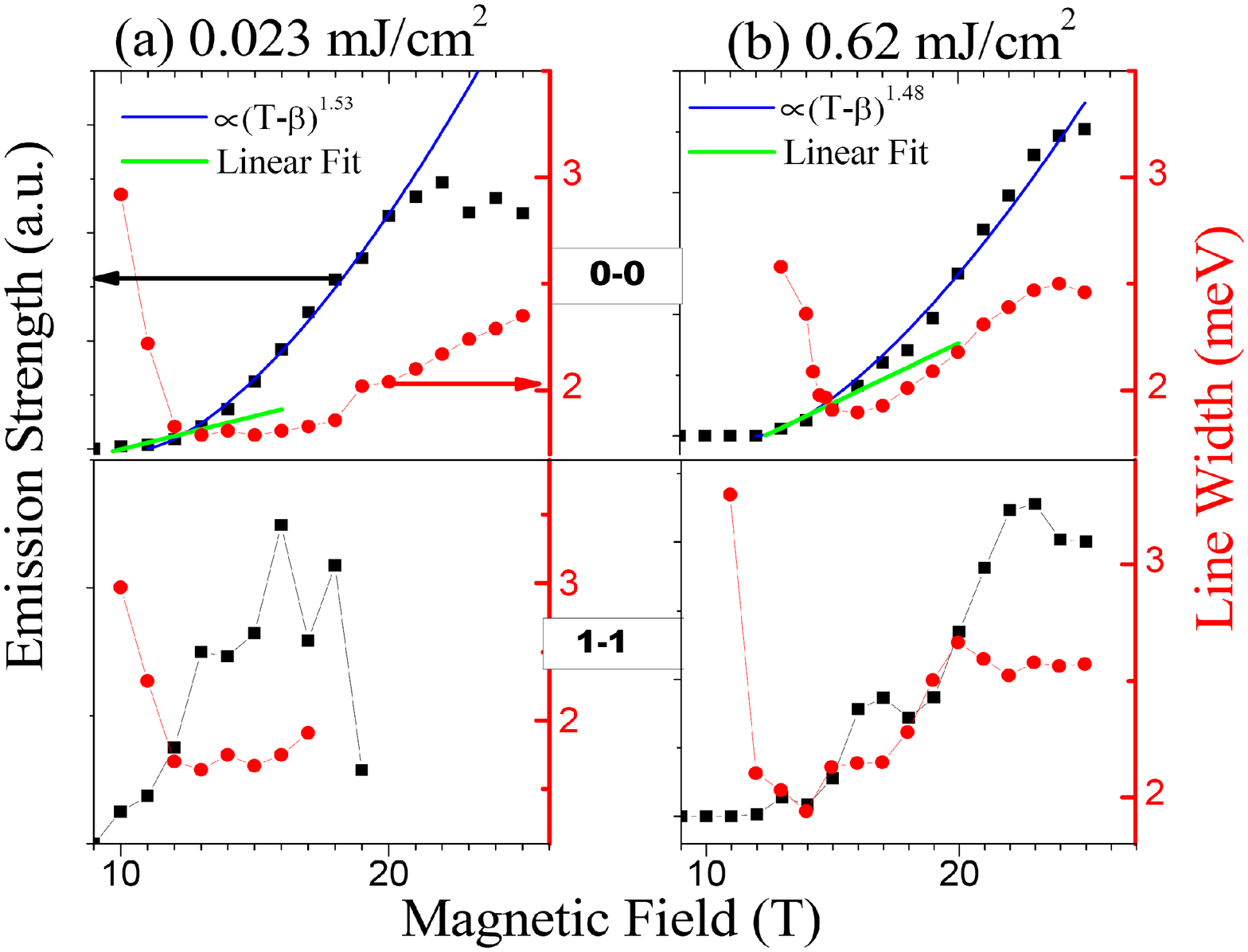} \caption{(Color Online) Emission strength
(black squares) and linewidth (red circles) of the narrow peak
from 0-0 and 1-1 LLs versus $B$ for different pump fluence of (a)
0.24 mJ/cm$^2$ and (b) 0.62 mJ/cm$^2$. The blue and green lines
represent the superlinear and linear scaling with $B$,
respectively.} \label{field scaling}
\end{figure}

To analyze the growth of the edge-collected signal with increasing
$B$, the integrated strength from individual LLs was fitted using
a Lorentzian lineshape analysis for the narrow blueshifted feature
and a Gaussian lineshape for the broad low energy redshifted
feature. A Gaussian lineshape is typical for spontaneous emission
while a Lorentzian lineshape originates from homogeneously
broadened systems. From the lineshape analysis, we obtained the
integrated emission signal (black squares) and linewidth (red
circles) traces in Fig. \ref{field scaling} for the pump fluence
of (a) 0.023 mJ/cm$^2$ and (b) 0.62 mJ/cm$^2$. At the lower
fluence (Fig. \ref{field scaling}(a)), the narrow emission 0-0 LL
peak emerges at approximately 11 T, then increases as $\sim B^2$
until 20 T (indicated by dashed line). The linewidth undergoes a
dramatic reduction (from 3 to $\sim$ 2 meV) as the field increases
from 10 to 13 T, followed by a slow increase as the field is
further increased. (Unless noted, we will refer only to the narrow
emission feature in further discussions.) The 1-1 LL feature shows
similar behavior with respect to threshold magnetic field and
linewidth, although the integrated PL intensity grows linearly and
begins to decrease at the highest field. The superlinear increase
for the 0-0 emission in Fig. 2 (a) was accompanied by an sharp
emission decrease from 1-1 LL, which indicates carrier depletion
from higher to lower LL when the fluence of 0.023 mJ/cm$^2$ could
barely saturate 0-0 LL.


At the higher fluence (Fig. \ref{field scaling}(b)), the emission
strength and linewidth display qualitatively similar behavior,
however, the 1-1 LL emission oscillates substantially as a
function of field. In addition, the increase in fluence allows for
the emission from 0-0 and 1-1 LLs over the entire field range of
25 T. Below 11 T, narrow emission is not observed for any LL. In
the case of 0-0 LL, the signal grows linearly with $B$ from 12-14
T(green lines) and the linewidth reveals a significant correlation
with the emission strength data.  In the linear regime, the
linewidth decreases monotonically versus $B$ until the emission
becomes superlinear, where the linewidth begins to increase.

\begin{figure}
\includegraphics[scale=0.45,trim=0 30 30 0,angle=0] {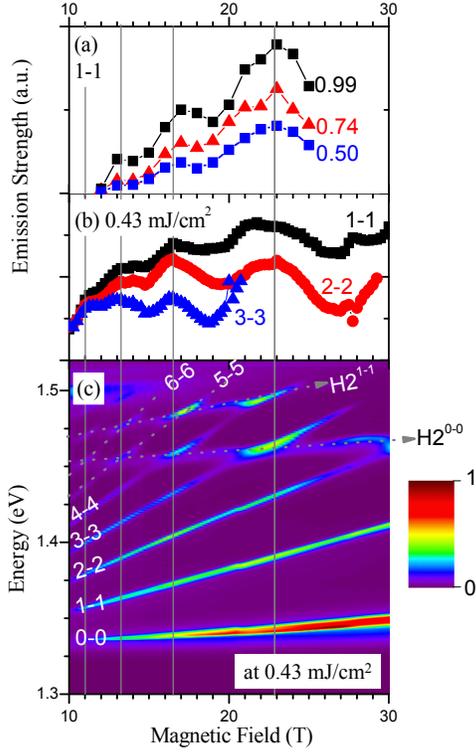}
\caption{(Color Online) Narrow line emission strength as a
function of magnetic field from 10 - 30 T for (a) the 1-1 LL at
various pump fluences and (b) 1-1, 2-2, and 3-3 LLs at a fixed
pump fluence of 0.43 mJ/cm$^2$ obtained from a lineshape analysis
of the emission data.  In (c), we present a contour plot of the
normalized emission strength versus field and energy for a fluence
of 0.43 mJ/cm$^2$.  The dashed gray lines in (c) show the mixing
points between various LLs and higher sublevels, and the vertical
lines show the correlation between the oscillation peaks and the
mixing energies.} \label{osc}
\end{figure}

To further investigate the origin of the oscillatory behavior with
field, in Figure \ref{osc} we plot the emission strength versus
the magnetic field for the 1-1 LL at different fluences (Fig.
\ref{osc}(a)) and for different LLs at a fixed fluence (Fig.
\ref{osc}(b)).  Above a fluence of 0.4 mJ/cm$^2$, the oscillations
are a universal feature, independent of any specific LL or
excitation fluence.  For comparative purposes, a contour map of
the field-dependent PL emission levels is presented in Fig.
\ref{osc}(c). In addition to the H1 LLs (labeled 0-0, 1-1, etc.),
the higher heavy hole states H2$^{0-0}$ and H2$^{1-1}$ are
observed (indicated by gray dashed lines) as reported in the
previous study\cite{Jho}. The gray vertical lines identify the
mixing energies and the oscillation peaks: a clear correlation is
seen between the mixing points of the H1-H2 levels and the peaks
in the oscillations of the emission.  At points where specific H1
and H2 LLs cross, a significant enhancement is seen in the
emission strength of those LLs.

\begin{figure}
\includegraphics[scale=0.45,trim=0 30 30 0,angle=0] {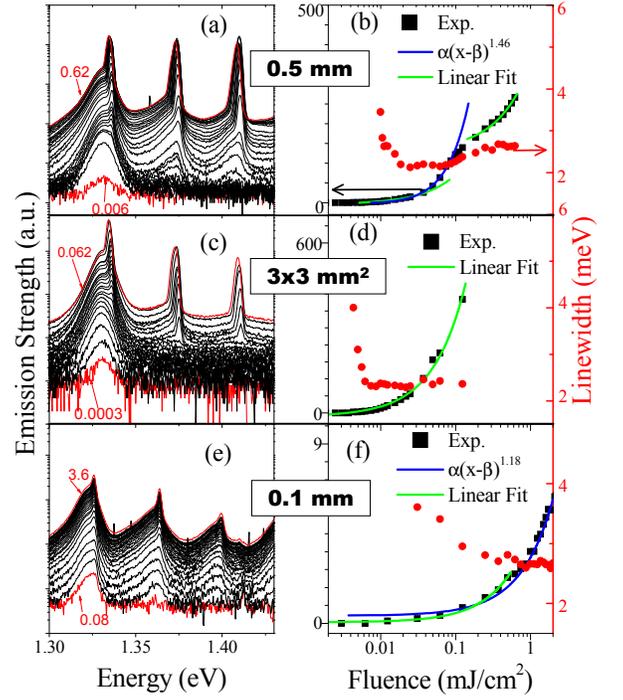}
\caption{(Color Online) Emission strength as a function of fluence
at fixed field at 20 T for different spot sizes of (a) 0.5 mm
diameter, (b) 3 mm and (c) 0.1 mm diamter. (d-f) Integrated
emission intensity (black squares) and linewidth (red circles) of
the narrow peak from the 0-0 LL versus fluence for different pump
spot sizes. The fluence (x-axis) is plotted on a logarithmic
scale. The inset of (a) shows the convolution method using a
Lorentzian for the sharp peak and a Gaussian for the broader
lower-energy peak. The blue (green) lines in (a-d) indicate the
superlinear (linear) regime.} \label{fluence}
\end{figure}

An observed spectral evolution of the PL emission as a function of
e-h density from a broad spontaneous emission to a narrow peak on
the blue side of the SE line, with subsequent further narrowing of
the peak followed by its broadening is consistent with the
scenario outlined in Sec. II. However, an observed spontaneous
emission line has a full width of $\sim$ 9 meV, larger than the
homogeneous linewidth of 5 meV implied in the above discussion.
The broad line may be caused by a variety of reasons, including
time evolution of the emission spectrum due to changing density of
carriers or spread of transition frequencies of e-h pairs.
Assuming the latter, for an inhomogeneously broadened line, we
calculated the numerically the growth rate and gain-induced index
change of electromagnetic waves for a Gaussian profile of
inhomogeneous broadening with full width $2 \gamma^*$. Simulations
show that the index change of 0.03 corresponding to the onset of
wave guiding for the same magnetic field of 12 T is achieved when
$2\gamma^* \simeq 6$ meV and $2\gamma \simeq 3$ meV, i.e. it
requires longer $T_2$. In this case the total SE linewidth is
$\simeq 9$ meV and the maximum index is reached at the blueshift
$\simeq 5$ meV from the central frequency.

After the wave guiding is reached, the process enters a saturated
SL regime accompanied by a line narrowing as a result of high-gain
amplification. A diminishing linewidth is expected as the spectral
components closest to the maximum of the gain spectrum get
amplified stronger than components with greater detuning. An
observed FWHM of the narrow peak changes with increasing e-h
density from $2\gamma \simeq 5$ meV to $2 \Delta \omega \simeq 2$
meV. If the maximum gain Eq.~(\ref{lase}) at the center of the
gain spectrum is equal to $g_0$, it is easy to obtain the
linewidth
\begin{equation} \label{narrow}
\frac{\Delta \omega}{\gamma} = \left[\frac{g_0 \mu/c}{{\rm ln
}[(G_0+1)/2]}-1 \right]^{1/2},
\end{equation}
  Where $G_0 = \exp{g_0 L \mu/c}$. For $L = 0.5$ mm (the pump spot size),
the observed narrowing by a factor of $\sim$ 2.5 is consistent
with the observed net amplification by a factor of order 100.

The transition to SF is accompanied by shortening of the pulse
from $\tau_{ASE} \sim L\mu/c \approx 5$ ps to $\tau_{SF}\sim
4/\omega_c \leq 2 /\sqrt{\gamma \gamma^*}$ \textbf{$\leq$ 1 ps}.
This shortening of the pulse duration leads to the observed line
broadening when the pulse duration becomes smaller than the
inverse total bandwidth of the stimulated emission peak. If we
identify $2\hbar/\tau_{SF} \sim \hbar g_{SF}$ with the observed
emission full width of 2.5 meV at $B = 25$ T in Fig. 2(b), the
value of the cooperative frequency is in the range $7-9$ meV
depending on the type of broadening we assume. This value is
definitely higher than both $\hbar \gamma $ and $\hbar
\sqrt{\gamma \gamma^*}$. Thus, spectral behavior is highly
suggestive of a continuous evolution from a regime where ASE is
the primary emission mechanism to an SF-dominated regime.

The intensity of a single SF pulse from a 0-0 LL scales with
electron density $N_e$ or the magnetic field $B$ as $I_{SF} \sim
\hbar \omega \Delta N/\tau_{SF} \propto
N_e^{3/2}(\Gamma(N_e))^{1/2}$ or $B^{3/2}(\Gamma(B))^2$, where
$\Gamma$ is the overlap factor as introduced in Eq. (1).

assuming that the pump pulse fluence $J_p \propto \Delta N \propto
N_e$. However, we cannot directly probe this superlinear scaling
since the emission strength is integrated over 1s (1000 shots).
Assuming all emitted photons are collected, we should observe a
linear dependence of the integrated emission strength from $N_e$
or $B$, which contradicts the observed superlinear dependence in
the SF regime.

We interpret the superlinear scaling of time-integrated intensity
of SF from 0-0 Landau level as the development of multiple SF
pulses after each pumping shot. As the first SF pulse develops on
the 0-0 LL and leads to a fast depletion of e-h pairs on this
level on a sub-picosecond scale, e-h pairs on the 1-1 LL do not
have time to recombine via a much slower ASE process. They lose
energy relaxing to the 0th level and emit another SF pulse,
leading to superlinear increase in the total fluence from 0th
level.

There are several lines of evidence that suggest that. First, the
superlinear scaling is observed only for the 0.5 mm spot,
approximately the SF coherence length $L_c$ as discussed in more
detail above and in the next section; see Fig. 4. SF cannot
develop for spot sizes smaller than the coherence length, whereas
for larger spots many independent uncorrelated SF pulses are
created, with the total emission resembling ASE rather than SF.
Furthermore, the superlinear increase in the 0-0 LL emission is
accompanied by a decrease in emission from higher LLs below 0.25
mJ/cm$^2$. Finally, the statistics of counts received by two edges
in Fig. 6 is consistent with the formation of two pulses from the
0-0 LL, as explained in Section IV C.

At very high pump fluences, SF may develop from two or three LLs
simultaneously, and the growth of the emission strength with the
pump fluence and magnetic field should slow down.

\subsection{Dependence of Photoluminescence on Excitation Area and Geometry}

The characteristics of the emission can be influenced by the
excitation spot size and by the geometry of the excitation region.
A spot size that is comparable to the SF coherence length $L_c$ is
optimal for SF emission.  A significantly larger spot size will
excite densities over a large area result in a number of
independently  SF-emitting (and uncorrelated) regions of dimension
on the order of $L_c$, while a smaller spot size will suppress the
formation of SF. Figure \ref{fluence} examines how the integrated
signal and linewidth in the narrow emission peak varies with
excitation area.   For these data, the field was held fixed at 20
T and the fluence was varied from 0.006 to 2 mJ/cm$^2$. The left
hand column displays the spectra (plotted logarithmically); the
right hand column plots the integrated intensity and linewidth
scaling. Figure \ref{fluence}(a,b) were obtained with spot size of
0.5 mm diameter, approximately equal to $L_c$ for our conditions.
In Fig. \ref{fluence}(b), we observe scaling behavior similar to
field scaling of 0-0 LL as in Fig. \ref{field scaling}(b), the
integrated strength (black squares) evolves from a threshold of
0.01 mJ/cm$^2$ through a linear regime (green line; 0.01-0.03
mJ/cm$^2$) to a superlinear regime (blue line; 0.03-0.2
mJ/cm$^2$). Above 0.2 mJ/cm$^2$ in Fig. 3(b), the signal resumes a
linear scaling after the SF from 0-0 LL is saturated as the level
is fully filled. The sustained linear regime in the 0-0 LL is
possibly associated with additional carriers passed on from higher
LL in the later stage after the SF pulse burst. The carrier
density which produces the linear scaling regime after saturation
can only be sufficient for ASE but not for SF. Such carrier
depletion from higher to lower LL was also manifested in Fig. 2
(a) via anti-correlation between 0-0 and 1-1 LL above 20 T. When
the laser spot was increased (decreased) to 3 mm (0.1 mm diameter)
as shown in Fig. \ref{fluence}(c,d) (Fig. \ref{fluence}(e,f)),
qualitatively different behavior is seen. Narrow emission peaks
were observed from each LL, but both the integrated signal (black
squares) and the linewidth (red circles) exhibits linear or nearly
linear scaling, and in both of these cases the linewidth
monotonically decreases with increasing fluence.

\begin{figure}
\includegraphics[scale=0.36,trim=20 20 20
0,angle=-90] {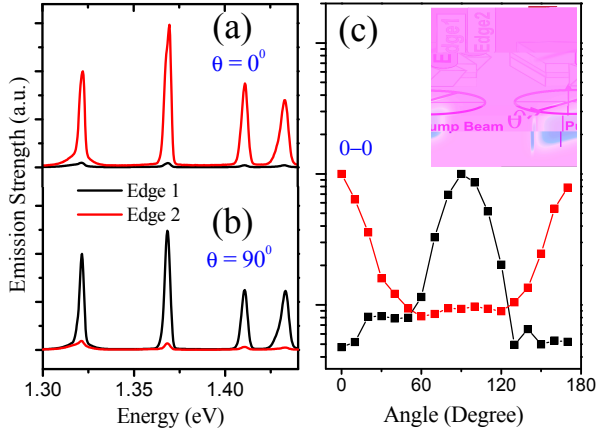} \caption{(Color Online) Edge emission
spectra measured from two orthogonally aligned fibers for the
angle $\theta$ at a 0$^{\circ}$  and b 90$^{\circ}$, where
$\theta$ is the angular separation between the longer beam axis
and the direction of edge 2 fiber as shown in the inset figure of
c. In c, the emission strength of 0-0 LL is plotted for edge 1
(black) and edge 2 (red) as a function of angle. }
\label{direction control}
\end{figure}

The shape of the excitation region can also be used to influences
the emission characteristics.  In particular, it is possible to
control the SF emission orientation through tailoring the geometry
of the gain region.  Using a cylindrical lens to generate a
rod-like 3 mm x 0.5 mm excitation region, Fig. \ref{direction
control} shows the output as a function of angle $\theta$ from
0$^{\circ}$ to 180$^{\circ}$ (see inset of Figure \ref{direction
control}(c)) for a fluence $F_{laser} =0.02~mJ/cm^2$ and $B$ = 25
T.  The maximum intensity follows the orientation of the long axis
of the gain region.  Tracing the output intensity displayed in
Fig. \ref{direction control}(c) for edges 1 (black) and 2 (red),
the emission is highly directional with a
full-width-at-half-maximum of 40$^{\circ}$.  This value,
comparable to the acceptance angle of our measurement, indicates
the emission is highly directional. The variance in emitted power
by 20 times between two directions are quite reasonable, since the
gain should scale exponentially with propagation length,
e$^{1.5/0.5}\sim 20$. Thus, the SF emission direction can be
manipulated through tailoring of the gain geometry. However, this
is not the unique feature of SF. The same behavior is expected
from high-gain ASE. In addition, the increased signal of $\sim$20
along the long axis relative to the short axis is due to
exponential gain, as expected for a stimulated process.

\begin{figure}\begin{center}
\includegraphics[scale=0.45,trim=20 20 20
30,angle=0] {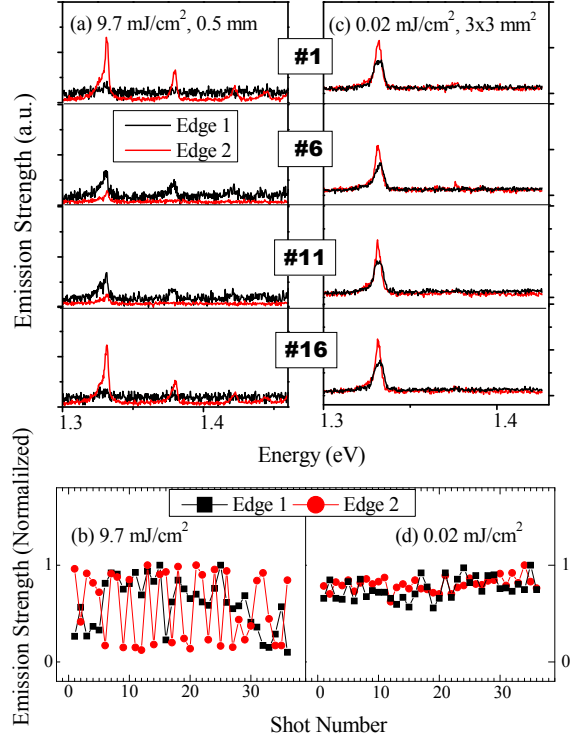}\end{center} \caption{(Color Online) Four
representative emission spectra at 25 T from edge 1 (black) and
edge 2 (red) fibers for (a) SF regime (9.7 mJ/cm$^2$ with 0.5 mm
spot size) and for (b) ASE regime (fluence of 0.02 mJ/cm$^2$ and
3x3 mm$^2$ spot size), excited from single laser pluse and
measured simultaneously. Normalized emission strength from the 0th
LL versus shot number in the (c) SF regime and the (d) ASE
regime.} \label{directionality}
\end{figure}

\subsection{Stochastic Emission Characteristics}

The strongest evidence for SF at high pumping fluences and
magnetic fields in our samples comes from the random
directionality experiment in the saturated SF regime ($>$ 0.2
mJ/cm$^2$). In this experiment, the pumped spot size was equal to
one coherence length 0.5 mm, and radiation has been collected
after single pulse excitation from two perpendicular edges
simultaneously. Since the pumped area is a circle, both SE and ASE
are emitted in all directions with the same intensity. This,
however, is no longer true for SF. In the development of SF,
initial quantum fluctuations grow to a macroscopic level and lead
to strong fluctuations of the delay time from pulse to pulse.
Initially, all propagation directions are equivalent. However,
after one pulse has propagated in one direction, all e-h pairs are
consumed along its path. Therefore, formation of the second pulse
in the direction traversing the path of the first pulse becomes
suppressed. Indeed, a SF pulse consumes virtually all e-h pairs
along its path leaving a narrow (of a few $\mu$m width) unpumped
stripe with negative inversion $\Delta N$. This stripe is {\it
anti-guiding} for the blueshifted SF emission. This means that the
SF radiation from any delayed pulse crossing the depleted stripe
will spread over the whole thickness of the sample after just one
wavelength $\lambda/\mu \sim 0.3$ $\mu$m of propagation length.
This reduces the overlap with an MQW layer to $\Gamma \sim
10^{-4}$, thus effectively quenching the pulse. Therefore, we
should observe strong anti-correlation between the emission
strengths received by the two edges after each pumping shot.

In order to verify this, we reduced the excitation pulse
repetition rate down to 20 Hz and collected emission from two
perpendicular edges whose outputs are mapped into spatially
separate images in CCD detector.

Using a pump fluence of 9.7 mJ/cm$^2$ at 25 T, Fig. 4(a)
illustrates typical spectra of SF after exciting one pump pulse,
collected through edge 1 (black) and edge 2 (red) fibers for four
representative shots, chosen from 36 shots in total.  In Fig.
4(b), the outputs from edge 1 and edge 2 fibers were normalized to
have the same maximum, although actual emission strength from edge
1 fiber was weaker, due to differences in the collection
efficiency of each fiber.

We clearly observe that the two fiber outputs are anticorrelated
over 19 different shots, which is about half of the total 36
shots. 16 events are those in which both edges receive significant
radiation. There is only one event when both edges receive little
radiation simultaneously. Over all of the shots, the maximum
observed emission strength in Fig. 4(b) fluctuates as much as
eight times the minimum value.  This is far greater than the pump
pulse fluctuation, on the order of a few percent, implying that
each SF burst is very directional and randomly changing from pulse
to pulse. The data can be best explained if there are two
consecutive SF pulses that can be formed per each pump shot,
according to the scenario suggested above. Each pulse can go
either to edge 1 or to edge 2. Of course, there are also pulses
generated in the opposite direction that we do not detect. At the
same time, our collecting prisms are very wide (1x1 mm$^2$), so we
collect most of the radiation propagating toward each of the two
edges. Then in 50 \% events both edges receive a SF pulse, or at
least some radiation if there is not enough electrons for the
second SF pulse and this second pulse is actually an ASE. In
another 50 \% events only one edge will receive both pulses.

At a lower $F_{\rm laser}$~=~0.02 mJ/cm$^2$ (obtained with a 3x3
mm$^2$ spot), qualitatively different behavior is seen;
Fig.~\ref{directionality}(b,d) shows omnidirectional emission on
every shot, as expected for ASE or SE. At this much lower power
with spot size much larger than the $L_c$, 1-1 LL is barely seen
while higher LLs are not occupied.

\subsection{Dependence of Photoluminescence on Temperature}

\begin{figure}
\includegraphics[scale=0.38,trim=0 20 100
0,angle=-90]{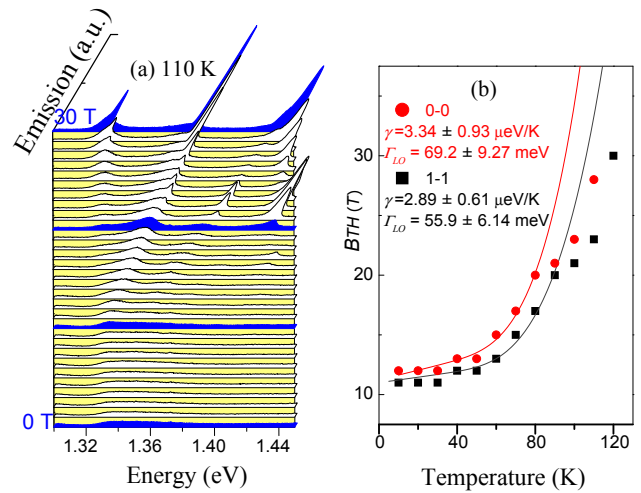} \caption{(Color Online) Emission spectra at
T = 110 K (b) The threshold field $B_{TH}(T)$ versus temperature
($T$) for the 0-0 LL (red circles) and for the 1-1 LL (black
squares) sharp emission features as a function of temperature. The
solid lines are fittings based on acoustic and optical phonon
contributions from Eq. (2).} \label{temp}
\end{figure}

Because the formation of SF inherently relies on the establishment
of a macroscopic coherence among the photoexcited carrier density,
we expect that the emission threshold and strength should depend
sensitively on temperature.  An increasing phonon number with
temperature should increases the intercarrier dephasing rate
$T_{2}$ and thus increase the threshold magnetic field and carrier
density.  In Fig. \ref{temp} (a), we display the emission spectra
obtained at a temperature of 110 K as a function of magnetic
field. We observe sharp features from the 0th LL at a much higher
magnetic field at $\sim$ 23T. In addition, the sharp features tend
to appear from higher LLs at lower fields with respect to the 0-0
LL as the field is increased. The sharp features, shortly appeared
from 2-2 and 3-3 LLs in the 21-23 and 26-30 T range, is possibly
associated with additional gain from increased density of state
when overlapped with e--light-hole (for 2-2 LL) or with second
confined level of e--heavy-hole (for 3-3 LL) transitions whose
energy states are slightly renormalized compared to the case of
low-power CW illuminations \cite{Jho}. As we follow the threshold
field $B_{TH}(T)$ for 0-0 LL (red circles) and for 1-1 LL (black
squares) sharp feature in Fig. 5 (b), we observe it is more
rapidly increasing above 50 K. Here, we define $B_{TH}$ as the
critical field for the appearance of narrow linewidth emission.
Even though the $B_{SF}$ is not coincident with $B_{TH}$ since the
sharp emission feature is generated by both the ASE and SF, we
will presume their temperature variation is similar. First, note
that the critical magnetic field $B_{SF}$ for SF generation is
obtained from relation between $\omega_c$ ($\propto \sqrt{N}$) and
the dephasing rate 2/$T_2$ \cite{Belyanin98,JhoSF}, $\omega_c >
2/T_2$. The LLs density of states is given by $N = eB/h$, i.e.,
proportional to $B$ and therefore to $\omega_c^2$. For
electron-phonon scattering, the dephasing rate 2/$T_2$ can be
expressed phenomenologically as:

\begin{equation}
2/T_2 \propto \Gamma_0 + \gamma T +
\Gamma_LO/[exp(E_{LO}/k_BT)-1],
\end{equation}

where $\Gamma_0$ is the width due to the inhomogeneous broadening
and $\gamma$ ($\Gamma_{LO}$) is fitting parameter which measures
the interaction with acoustic phonon (polar LO phonons).  Thus,
writing $eB_{SF}/h \propto \omega_c^2 =(2/T_2)^2$, we find

\begin{equation}
B_{SF} \propto (\Gamma_0 + \gamma T +
\Gamma_LO/[exp(E_{LO}/k_BT)-1])^2,
\end{equation}

Since the LO phonon energy $E_{LO}$ in our sample structure, which
is expected to be similar to that of GaAs-based QW ($\sim$ 36
meV), is much larger than thermal energy ($k_{B}T$) in our
temperature range, we tentatively identify the acoustic phonon
contribution as a dominant temperature mechanism for varying
$B_{SF}$.  The  red (black) curve in Fig. 7(b) is the fitting
based on Eq. (9) for 0-0 LL (1-1 LL), where we can obtain the
comparative values of $\gamma/\Gamma_0$ and
$\Gamma_{LO}/\Gamma_0$, respectively. When $\Gamma_0$ being
assumed to be the same with the minimum line-width obtained from
Fig. 2(b) (=1.9 meV), $\gamma$ ($\Gamma_{LO}$) is smaller (larger)
than that of 2-dimensional exciton case by 2-3 times (3-4 times)
\cite{Zhao} while being very similar to the zero-dimensional
case.\cite{Favero} The deviation of the fitting curve above 80 K
for 0-0 LL and above 90 K for 1-1 LL is possibly associated with
carrier delocalization and/or deionized impurities at high
temperatures.\cite{Lee} $\Gamma_0$ ($\sim$1.9 meV) indicates the
SF pulse duration is shorter than 700 fs (=2$\hbar/\Gamma_0$)
while the temperature dependence of $B_{TH}(T)$ agrees with the
dephasing dynamics in zero-dimensional system.

\subsection{Emission characteristics with stretched excitation pulse width}

\begin{figure}\begin{center}
\includegraphics[scale=0.4,trim=50 50 50
0,angle=0]{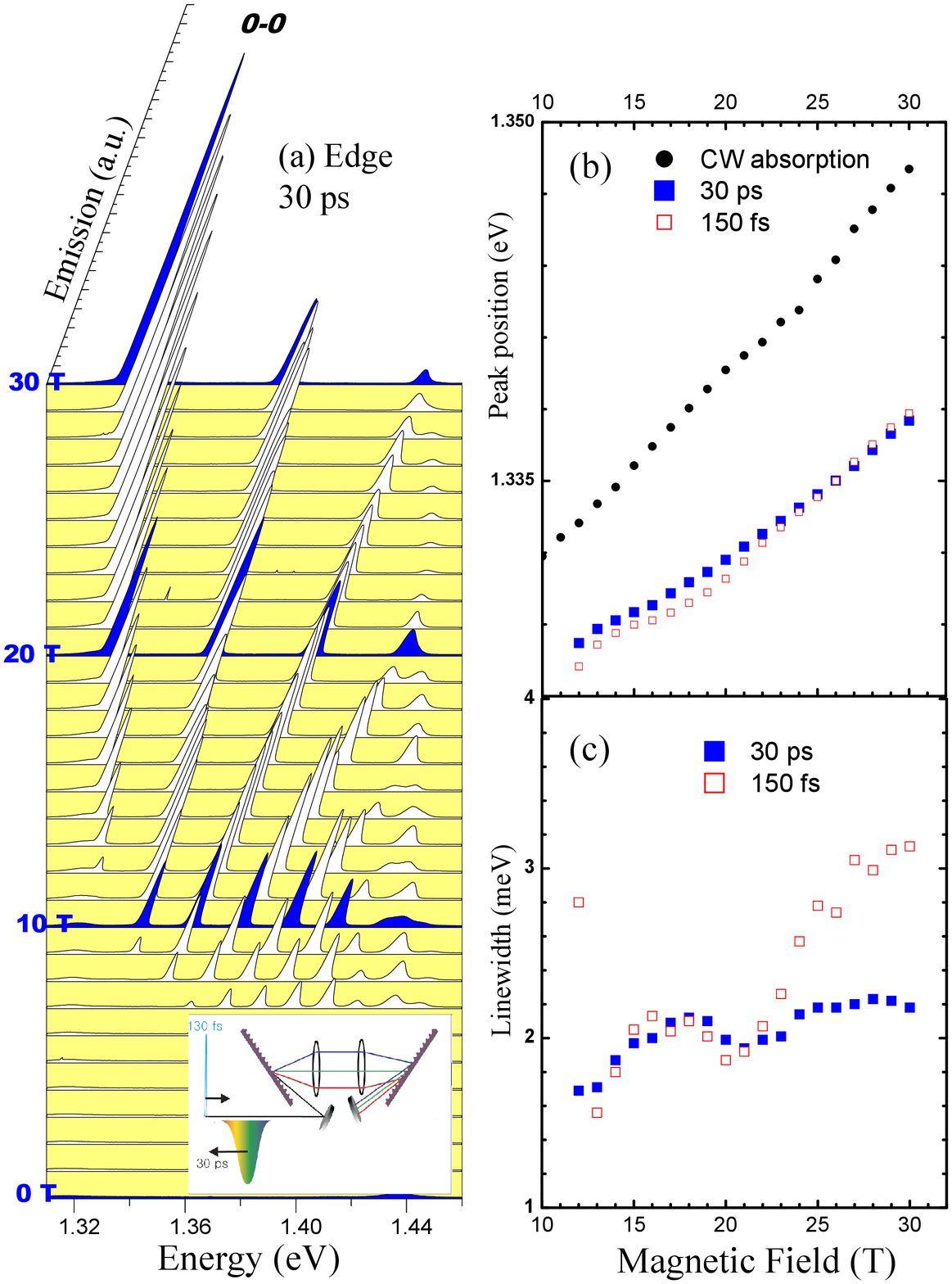}\end{center} \caption{(Color Online) (a)
Emission spectra as a function of magnetic field for the
excitation pulse width of 30 ps. The inset shows the pulse
stretching scheme using two wavelength dispersing diffraction
gratings. (b) 0-0 LL peak energy positions for the pulse width of
30 ps (solid square) and 150 fs (open square), compared with low
power CW absorption. (c) Linewidth of the sharp feature from 0-0
LL versus $B$ for 30 ps (solid square) and for 150 fs (open
square).} \label{pulse}
\end{figure}

Finally, we investigate how the emissions depend on the excitation
pulse width, as being compared with the delay time $t_d$. For the
optimal coherence length $L_c$, $t_d$ is roughly estimated around
10 ps when the pumping fluence of $\sim 0.1$ mJ/$cm^{2}$ almost
saturates the available states in 0-0 LL (cf., Fig.
4(b)).\cite{Belyanin03} In this regard, in figure 8., the
characteristics of the emission were examined with pulsewidth
broadened up to 30 ps. The pulse stretcher was composed by a
grating pair, a lens pair, and mirrors as cartooned in the inset
of Fig. 8(a). The sample temperature and the laser fluence were
fixed at 10 K and at around 0.1 mJ/cm$^{2}$ with which total
excitation carrier density can reach $\sim$10$^{13}$ /cm$^{2}$.

Fig. 8(a) shows corresponding emission spectra in the edge
collection geometry with magnetic field from 0 to 30 T. The 1-1 LL
emission strength, compared with 0-0 LL, was relatively much
stronger than that of Fig. 1 (b). This indicates a certain amount
of slow carrier relaxation from higher LL down to 0-0 LL exists in
a temporal scale shorter than 30 ps. We also observe the evident
oscillatory behavior at 1-1 and higher LLs which was introduced in
Fig.3 with 150 fs pulsewidth excitation. In Fig. 8(b) we traced
the peak energy positions of sharp Lorentzian feature in 0-0 LL at
different excitation pulsewidths which were redshifted from
previously studied case of Gaussian-fitted CW
absorption\cite{Jho}. The red-shifted peak position for pulsed
excitation indicates not only the existence of localized states in
these strained samples\cite{Jho} but also the band gap
renormalization \cite{Butov3,Tatham}. We also note the effect of
bandgap renormalization was stronger at 150 fs pulse but the
difference between 150 fs and 30 ps was almost suppressed above 25
T. The 150 fs pulse creates carriers more abruptly than 30 ps
pulse and the following carrier-carrier interactions will be
enhanced to screen the excitonic Coulomb interactions more
effectively. The disappearance of the peak energy separation
between 150 fs and 30 ps above 25 T, therefore, could be possibly
associated with the combined effects of increased carrier density
and reduced screening effects because of more strongly confined
wavefunctions via $B$ increment.  The slope of peak position
versus $B$ in Fig. 8 (b) became more similar to CW absorption case
with the increasing field, which suggests the reduction of
effective mass renormalization in spite of increased carrier
density at higher fields. More details on the field dependent
variation of bandgap and effective mass renormalization will be
discussed elsewhere.

To further characterize the emission in Fig. 8(c), we measured the
linewidths for 150 fs and for 30 ps for Lorentzian sharp features
in 0-0 LL. Fig. 8(c) displays the clear trend of increasing
linewidths with $B$ for 150 fs in contrast to those for 30 ps. As
introduced earlier, 30 ps is much longer than the estimated delay
time $t_d$ to build coherence within $L_c$ in our system and the
intra-relaxation time on the order of 500 fs\cite{Tatham}. Since
the increment of linewidth with $B$ was a signature of SF, plateau
in the linewidth trace for 30 ps in Fig. 8(c) implies the
saturated amplified spontaneous emission, not reaching SF regime
yet.

\section{Conclusion}\label{conc}
We have performed magneto-photoluminescence measurements in
In$_x$Ga$_{1-x}$As multiple quantum wells in high magnetic fields
using intense femtosecond pulses. The resulting density and energy
confinement is sufficient to generate a spontaneous macroscopic
polarization that decays through the emission of SF pulses.  Our
experiments demonstrate multiple evidence for the cooperative SF
regime of spontaneous recombination as revealed by the spectral
behavior, magnetic field and fluence scalings, spot size
dependence, spatial and temporal correlations. We further probe
the appropriate conditions for observing SF by exploiting its
spectral features and dependence on the temperature and focal
geometry.

\section*{Acknowledgments}

This work was supported by the Korean Research Foundation Grant
(MOEHRD, KRF-2007-313-C00218), the NSF ITR program (DMR-032547)
and the Bio-imaging Research Center at GIST. A portion of this
work was performed at the National High Magnetic Field Laboratory,
supported by NSF Cooperative Agreement No. DMR-0084173 and by the
State of Florida. A.B. acknowledges the support from the NSF
CAREER program (0547019) and NSF PIRE and ERC programs.


\begin{thebibliography}{26}
\bibitem{rabi} $\Omega_R$ also contains non-resonant ''local
field'' corrections from screened Coulomb interactions. We will
however neglect them here, assuming $\Omega_R = dA/2\hbar$, where
the dipole moment $d/e \simeq 0.5$ nm. The local field effect can
be important for SF of a dense small sample of size much smaller
than the wavelength. In our case it creates only a small
correction to the SF dynamics.
\end{thebibliography}
\end{document}